\documentclass[aps,showpacs,twocolumn,amsmath,amssymb,amsfonts,eqsecnum]{revtex4-1}
\usepackage{graphicx}
\usepackage{color}
\usepackage{natbib}
\usepackage{epstopdf}
\usepackage{hyperref}

\newcommand{\sign}{\textup{sign}}
\newcommand\eq[1]{Eq.~(\ref{#1})}

\begin{document}
\title{Exact solution of Ginzburg's $\Psi$-theory for the Casimir force in $^4$He superfluid film}
\author{Daniel Dantchev$^{1,2}$\thanks{e-mail:
		daniel@imbm.bas.bg}, Joseph Rudnick$^{1}$\thanks{e-mail:
		jrudnick@physics.ucla.edu}, Vassil Vassilev$^{2}$\thanks{e-mail:vasilvas@imbm.bas.bg} and Peter Djondjorov$^{2}$\thanks{e-mail:padjon@imbm.bas.bg}} \affiliation{ $^1$ Department of Physics and Astronomy, UCLA, Los Angeles,
	California 90095-1547, USA,\\$^2$Institute of
	Mechanics, Bulgarian Academy of Sciences, Academic Georgy Bonchev St. building 4,
	1113 Sofia, Bulgaria
}

\date{\today}

\begin{abstract}
We present an analytical solution of the Ginzburg's $\Psi$-theory for the behavior of the Casimir force in a film of $^4$He in equilibrium with its vapor near the superfluid transition point, and we revisit the corresponding experiments  \cite{GC99} and \cite{GSGC2006} in terms of our findings. We find reasonably good agreement between the $\Psi$-theory predictions and the experimental data. Our calculated force is attractive, and the largest absolute value of the scaling function is $1.848$, while experiment yields $1.30$. The position of the extremum is predicted to be at $x=(L/\xi_0)(T/T_\lambda-1)^{1/\nu}=\pi$, while experiment is consistent with $x=3.8$. Here $L$ is the thickness of the film, $T_\lambda$ is the bulk critical temperature and $\xi_0$ is the correlation length amplitude  of the system for $T>T_\lambda$. 

\end{abstract}

\maketitle

\section{Introduction}

\subsection{Critical Casimir force near the $\lambda$ transition in $^4$He}

It is now a well established experimental fact \cite{GC99,GSGC2006} that the thickness of a helium film in equilibrium with its vapor decreases near and below the bulk transition into a superfluid state. The phenomenon has been discussed theoretically in a series of works; see, e.g., Refs. \cite{KD92a,KD92b,ZRK2004,DKD2005,MGD2007,ZSRKC2007,MGD2007,H2007,D2013,V2015}. Among the methods used are renormalization group techniques \cite{KD92a,KD92b}, mean-field type theories \cite{ZSRKC2007,MGD2007} and Monte Carlo calculations \cite{DKD2005,H2007,V2015}. It should be noted that in all of the above approaches it is assumed that the microscopic molecular interactions are not altered by the transition and that the observed change in the thickness thus results from the  cooperative behavior of the constituents of the fluid system. Furthermore, the overall behavior of the force is in a relatively good agreement with finite-size critical point scaling theory \cite{Ba83,C88,Ped90,PE90,BDT2000}.   

An inspection of the range of theoretical approaches used to study the Casimir force in helium films reveals that the problem has, so far, not been studied in the context of the so-called $\Psi$-theory of Ginzburg and co-authors \cite{GS76,GS82}. This theory has been used by Ginzburg, et. al., to describe a variety of phenomena observed in Helium films and represents a portion of the research on Helium for which  Ginzburg was recently awarded the Nobel prize in physics. In the current study we aim to fill that gap by applying  $\Psi$ theory to calculate the critical Casimir force of a helium film that is subject only to short-ranged interactions, which is to say we neglect the van der Waals interaction between the film and its substrate. 

We study the Casimir force in a horizontally positioned liquid $^4$He film supported on a substrate when that film is in equilibrium with its vapor. We will do this for temperatures at, and close to, the critical temperature, $T_\lambda$, of $^4$He at its bulk phase transition from a normal to a superfluid state. 

\subsection{Some data and facts from the experiment}

The continuous phase transition in $^4$He from a normal to a superfluid state, referred to as the $\lambda$ transition because of the temperature dependence of the specific heat,  occurs  at a temperature \cite{GS76} $T_\lambda=2.172$ $^\circ$K at a saturated-vapor pressure $p_\lambda =0.05$ atm and density \cite{W67,GS82} $\rho_\lambda=0.1459$ g/cm$^3$. We note that while the density changes continuously through the transition its temperature gradient varies discontinuously \cite{W67,DB98}. 

The critical exponents of systems, that belong to the $O(2)$ universality class of $O(n)$, $n\ge 2$ of systems with continuous symmetry of the order parameter, are \cite{KS2001,ZJ2002}
\begin{eqnarray}
\label{eq:crit_exp}
\alpha &=& -0.011\pm 0.004  , \nu= 0.6703 \pm 0.0013,\nonumber \\
\eta &= & 0.0354\pm 0.0025.
\end{eqnarray}
Since hyperscaling holds, all critical exponents can be determined from, say, $\nu$ and $\eta$ using the appropriate scaling relations. 

\subsection{The Casimir force}

Finite-size scaling theory \cite{Ba83,C88,Ped90,PE90,BDT2000} for systems in which hyperscaling holds predicts a Casimir force of a system with a film geometry $\infty^{d-1}\times L$ of the form 
\begin{equation}
\label{eq:Cas_force_def}
F_{\rm Cas}(T,L)=L^{-d} X_{\rm Cas}(a_t \hat{t} L^{1/\nu}).
\end{equation}
Here $X_{\rm Cas}$ is a universal scaling function that depends on the bulk and surface universality classes, $\hat{t}=(T-T_\lambda)/T_\lambda$, and $a_t$ is a nonuniversal metric factor. Helium 4 belongs to the $O(2)$ bulk universality class and the boundary conditions on a helium film on a solid substrate that is in equilibrium with vapor are Dirichlet, in that the superfluid order parameter vanishes at the boundaries. 

The Casimir force $F_{\rm Cas}(T,L)$ can be expressed in terms of an excess pressure $P_L(T)-P_b(T)$:
\begin{equation} \label{Casimir}
F_{\rm Cas}(T,L)= P_L(T)-P_b(T)
\end{equation} 
Here $P_L$ is the pressure on the finite system, while $P_b$ is the pressure in the infinite system. The Casimir force results from finite size effects, which are especially pronounced and of {\it universal} character near a critical point of the system. The above definition is equivalent to another, commonly used, relationship \cite{E90book,K94,BDT2000}
\begin{equation}
\label{grand_can}
F_{\rm Cas}(T,L)\equiv-\frac{\partial\omega_{\rm ex}(T,L)}{\partial L}=-\frac{\partial\omega_L(T,L)}{\partial L}-P_b,
\end{equation}
where $\omega_{\rm ex}=\omega_L-L\,\omega_b$ is the excess grand potential per unit area, $\omega_L$ being the grand canonical potential of the finite system, again per unit area, and $\omega_b$ is the grand potential per unit volume of the infinite system. The equivalence between the definitions \eq{Casimir} and \eq{grand_can} arises from the observation that  $\omega_b=- P_b$ while for the finite system with surface area $A$ and thickness $L$ one has $\omega_L=\lim_{A\to \infty} \Omega_L/A$, with $-\partial\omega_L(T,L)/\partial L=P_L$. 

\subsection{On the Casimir force in a class of systems}

It is possible to derive a simple expression for the Casimir force in systems in which the order parameter is found by minimizing a potential that does not explicitly depend on the coordinate perpendicular to the film surface. For purposes of notation we denote the spatial coordinate perpendicular to the substrate by $z$.  We consider systems in which the grand potential per unit area ${\cal \omega_A}$ is obtained by minimizing the functional 
\begin{equation}
\label{eq:functional}
{\cal \omega_A}=\int_{0}^{L} {\cal L}\left[ \phi(z),\dot{\phi}(z)\right] dz,
\end{equation}
where $\phi(z)$ is the local value of the order parameter at coordinate $z$, and $\dot{\phi}(z)\equiv d\phi(z)/dz$. We take $\cal L$ to be of the form
\begin{equation}
\label{eq:L}
\mathcal{L}=\frac{1}{2} \dot{\phi}^2(z)-f[\phi(z)]. 
\end{equation}
Following Gelfand and  Fomin \cite[pp. 54-56]{GF63} it is easy to show that the functional derivative of  ${\cal \omega_A}$ with respect to the independent variable $z$ at $z=L$ is 
\begin{equation}
\label{eq:funct_deriv}
-\left(\frac{\delta {\cal \omega_A} }{\delta z}\right)\Bigg|_{z=L} =-\left(\dot{\phi} \frac{\partial {\cal L}}{\partial \dot{\phi}}-{\cal L}\right)\Bigg|_{z=L}.
\end{equation}
Taking into account the physical meaning of this functional derivative and performing the requisite calculations we obtain 
\begin{equation}
\label{eq:pl_def}
P_L\equiv \left(\frac{\delta {\cal \omega_A} }{\delta z}\right)\Bigg|_{z=L} =\left(\dfrac{1}{2}\dot{\phi}^2+f(\phi)\right)\Bigg|_{z=L}.
\end{equation}
The extrema of the functional ${\cal \omega_A}$ are determined by the solutions of the corresponding Euler-Lagrange equation
\begin{equation}
\label{eq:EL_eq}
\dfrac{d}{dz} \frac{\partial {\cal \omega_A}}{\partial \dot{\phi}}-\frac{\partial {\cal \omega_A}}{\partial \phi}=0,
\end{equation}
which leads to 
\begin{equation}
\label{eq:EL_eq_expl}
\dfrac{d}{dz} \dot{\phi}+\frac{\partial f}{\partial \phi}=0.
\end{equation}
Multiplying by $\dot{\phi}$ and integrating one obtains the corresponding first integral of the above second-order differential equation. The result is
\begin{equation}
\label{eq:first_int}
\dfrac{1}{2}\dot{\phi}^2+f(\phi)={\rm const}=P_{L}.
\end{equation}
Thus, the expression for $P_L$ has the same values at {\it any} point of the liquid film. 

Let now assume that the boundary conditions are such that there is a point at which $\dot{\phi}=0$ and let $\phi_0$ be the value of $\phi$ at that point. Then we arrive at the very simple expression for the pressure on the boundaries of  the finite system
\begin{equation}
\label{eq:PL_final}
P_L=f(\phi_0). 
\end{equation}
When the system is infinite the gradient term decreases with distance from a boundary, asymptoting to zero in the bulk  within the type of theories we consider.  It is easy to verify that the bulk pressure is 
\begin{equation}
\label{eq:Pb_final}
P_b=f(\phi_b), 
\end{equation}
where $\phi_b$ is the solution of the equation $\partial f/\partial \phi=0$, for which $\omega_b=-f(\phi)$ attains its minimum. The excess pressure,  and hence the Casimir force, is 
\begin{equation}
\label{eq:Cas_gen}
F_{\rm Cas}\equiv P_L-P_b=f(\phi_0)-f(\phi_b). 
\end{equation}
The above expression, as we will see, is very convenient for the determination of the Casimir force  in a system that can be described by a functional of the type given in \eq{eq:functional}. It has previously been used for systems described by the Ginzburg-Landau-Wilson functional \cite{ZSRKC2007,DVD2016,Note1}.

\section{The model}

We now consider a film with thickness $L$ of liquid $^4$He that is in equilibrium with its vapor. We suppose the film to be parallel to the $(x,y)$ plane and its thickness to be along the $z$ axis. A constituent of the liquid film with total density $\rho$ is in the superfluid state with density $\rho_s(z)$ while the other one with density $\rho_n(z)$ is in the normal state. Obviously
\begin{equation}
\label{eq:rhos}
\rho(z)=\rho_n(z)+\rho_s(z).
\end{equation} 
We consider two order parameters: a one-component order parameter $\rho_n$ to represent the normal fluid and two-component parameter $\Psi_s=\eta \exp(i \varphi)$ to stand for the superfluid portion of it. As usual, we take $\eta=\eta(z)$ and $\varphi=\varphi(z)$ to be real valued functions and, thus, $|\Psi_s|=\eta$ with the identification that
\begin{equation}
\label{eq:rhos_eta}
\rho_s=m |\Psi_s|^2=m\eta^2, 
\end{equation}
where $m$ is the mass of the helium atom.  A spatial gradient of the phase of the $\Psi_s$ function gives rise to the superfluid velocity via the relationship
\begin{equation}
\label{eq:vs}
\vec{v}_s=\frac{\hbar}{m}  \nabla \varphi. 
\end{equation}

In the remainder of this article we consider only the case of a fluid at rest. Then one can take $\Psi_s$ to be a real function characterized solely by its amplitude $\eta$.

In terms of $\rho_s=m|\Psi_s|^2$ and $\rho_n$, the total amount of helium atoms in the fluid (normalized per unite area) is
\begin{equation}
\label{eq:rho}
\rho \equiv \frac{1}{ L}\int_{0}^{L} \left[ \rho_s(z) + \rho_n(z)\right ] dz=\frac{1}{ L}\int_{0}^{L} \rho(z) dz,
\end{equation}
where the value of the overall average density $\rho$ is fixed by the chemical potential $\mu$. The above equation intertwines the profiles $\rho_s$ and $\rho_n$.  For $\rho_s$ the natural boundary conditions at both the substrate-fluid interface and the fluid - vapor interfaces are
\begin{equation}
\label{eq:bcpsis}
\rho_s(0)= \rho_s(L)=0 \Leftrightarrow \Psi_s(0)=\Psi_s(L)=0.
\end{equation}
The corresponding natural boundary conditions for $\rho_n$ depend on the interface. At the liquid-vapor interface one has 
\begin{equation}
\label{eq:bcpsin}
\rho_n(L)=\rho_b(T),
\end{equation}
where $\rho_b(T)$ is the bulk density of the liquid helium at temperature $T$; at the substrate-liquid interface one has the so-called ``dead" layers. In these layers $^4$He has solid-like properties, i.e., it does not possess a properties of a liquid,  and it is immobilized at the boundary. This implies that there is some sort of close packing of the helium atoms. The number of layers is generally small---from two well below $T_\lambda$ to the order of 10 in the vicinity of that temperature.  This can be thought of as a sort of adjusted thickness of the liquid films and will be ignored in our theory. Thus, we will assume that the boundary condition  \eqref{eq:bcpsin} is fulfilled at the both boundaries of the system, i.e., that
\begin{equation}
\label{eq:bcpsinormal}
\rho_{n}(0)=\rho_n(L)=\rho_b(T).
\end{equation}

Since we are addressing a spatially inhomogeneous problem, its  proper treatment requires the minimization of the total thermodynamic potential ${\cal \omega_A}(\mu,T)$ \cite{GS82,So73}, 
which is normalized per unit area, simultaneously with respect to $\Psi(z)$ and $\rho(z)$. Hereafter the dot will mean a differentiation with respect to the coordinate $z$.

\subsection*{A realization of the model within the so-called $\Psi$ theory} 

We take as our basic variables $\rho_s$ and $\rho$. We assume that they both vary within the film, so our system will depend on $\rho_s$ and $\rho$ and their gradients  $\dot{\rho_s}$ and $\dot{\rho}$. If however, the gradient of  $\rho$ is small, spatial derivatives
of $\rho$ can be neglected. For temperatures well below the liquid-vapor critical point we will take $\rho$ to be a constant within the film, i.e., $\rho$ is $z$-independent. This implies near the $\lambda$ point one can treat helium as an incompressible liquid. This is what is done in \cite{GS76} and \cite{So73}.

For the total thermodynamic potential ${\cal\omega_A}(\mu,T)$ per unit area one has
\begin{equation}
\label{eq:funct}
{\cal\omega_A}(\mu,T) =  \int_{-L/2}^{L/2} \left[\omega(\mu,T,\rho, \Psi_s,\dot{\Psi_s}) - \mu \rho \right] dz,
\end{equation}
where $\omega(z)$ is the local density of this potential per unit area. Here $\omega=\omega_I(\mu,T, \rho)+\omega_{II}(\mu,T,\Psi_s,\dot{\Psi_s})$, where $\omega_I$ is the local potential density of the normal fluid and $\omega_{II}$ is that of the superfluid. Since $\mu, T$ and $\rho$ are constants through the thickness of the film, one concludes that the terms $\omega_I$ and $\mu \rho$ will generate only bulk-like contributions, after the integration. For this reason we will not be interested in the specifics of these terms.   Following \cite{GS82}, one can write 
\begin{equation}
\label{eq:omegaII}
\omega_{II}=\omega_{II,0}+\frac{1}{2 m}|-i\hbar\dot{\Psi}_s|^2,
\end{equation}
where $\omega_{II,0}=\omega_{II,0}(\mu,T,|\Psi_s|^2)$ captures the corresponding bulk potential density of the infinite system. The gradient term can easily be rewritten in the equivalent forms
\begin{eqnarray}
\label{eq:eq_forms_grad_term}
|-i\hbar\dot{\Psi}_s|^2 &=& \frac{\hbar^2}{2m}\dot{\eta}^2+\frac{\hbar^2}{2m}\eta^2 \dot{\varphi}^2 \nonumber \\
&=& \frac{\hbar^2}{8m^2} \frac{\dot{\rho}_s^2}{\rho_s}+\frac{1}{2}\rho_s v_s^2.
\end{eqnarray}
For a fluid at rest $v_s=0$. 

In accord with \cite{GS82,GS76}, we take $\omega_{II,0}$ to be of the form
\begin{eqnarray}
\label{eq:II_pot}
\lefteqn{\omega_{II,0}(\mu,T,|\Psi_s|^2)=\dfrac{3T_\lambda \Delta C_\mu}{3+M}\left[-\tau |\tau|^{1/3} \left|\frac{\Psi_s}{\Psi_{s,e0}}\right |^2 \right. } \nonumber\\
&&\left. + \frac{1-M}{2}|\tau|^{2/3}\left|\frac{\Psi_s}{\Psi_{s,e0}}\right |^4 + \frac{M}{3}\left|\frac{\Psi_s}{\Psi_{s,e0}}\right |^6 \right], 
\end{eqnarray}
where $T_\lambda(\mu)$ is the $\lambda$-transition temperature in
equilibrium with saturated vapor, $T_\lambda(\rho_\lambda) = 2.172$ K, $\rho_\lambda = 0.146$ g cm$^{-3}$. Here
\begin{equation}
\label{eq:tau}
\tau=(T_\lambda(\mu)-T)/T_\lambda(\mu),
\end{equation}
 $\Delta C_\mu $ is the specific heat jump
at the $\lambda$ point $\Delta C_\mu = \Delta C_p = 0.76 \times 10^7$ erg cm$^{-3}$ K$^{-1}$, $M$ is a parameter of
the theory, and $\Psi_{s,e0}$ is the amplitude of the temperature dependence of
the equilibrium value of $\Psi_s$ in bulk helium,
\begin{equation}
\label{eq:psi_eq_value}
\Psi_{s,e}(\tau) =\Psi_{s,e0}\, \tau^\beta = 0.23 \times 
10^{12} \, \tau^{1/3} \, {\rm cm}^{-3/2}.
\end{equation}
The value of $\Psi_{s,e0}$ is, as usual \cite{GS82}, determined by the equation 
\begin{equation}
\label{eq:bulk_psi}
\left(\frac{\partial \omega_{II,0}}{\partial |\Psi|^2}\right)_{\mu,T}=0
\end{equation}
so as to be in accord with the experimental data 
\begin{eqnarray}
\label{eq:bul_value_Psi}
\rho_{se}&=& m|\Psi|^2= 1.43\, \rho_{\lambda}T_\lambda^{2/3} \tau^{2/3} \nonumber \\
&=& 0.35\,\tau^{2/3}\; {\rm g}\, {\rm cm}^{-3} =\rho_{s0}\, \tau^\zeta
\end{eqnarray}
with $\zeta\simeq 2\beta\simeq 2/3$. As is clear from \eq{eq:II_pot}, it is convenient to introduce the reduced variable
\begin{equation}
\label{eq:small_psi}
\psi=\frac{\Psi_s}{\Psi_{s,e0}}.
\end{equation}
Then, \eq{eq:II_pot} becomes 
\begin{eqnarray}
\label{eq:II_pot_small_psi}
\omega_{II,0}&=&\dfrac{3T_\lambda \Delta C_\mu}{3+M} \left[-\tau |\tau|^{1/3} \left|\psi\right|^2  \right. \\
&& \left. + \frac{1-M}{2} |\tau|^{2/3} \left|\psi\right|^4 + \frac{M}{3}\left|\psi\right|^6 \right]. \nonumber
\end{eqnarray}

The above expressions for $\omega$ and $\omega_{II}$ are consistent with a close approximation to the critical exponents in which $\alpha=0$, $\nu=2/3$ and the anomalous dimension exponent $\eta$ is zero.  They define an effective $3$-dimensional theory for the behavior of the helium films. The best known values of the critical exponents $\alpha$ and $\nu$ for helium are given above, in \eq{eq:crit_exp}. 

The conditions for the minimum of ${\cal\omega_A}(\mu,T)$ are given by the corresponding Euler-Lagrange equations (see also Eqs. (1.3) and (1.4) in \cite{So73}, or Eq. (3.41) in \cite{GS76}), which read
\begin{equation}
\label{eq:EL_rho}
-\frac{\partial \omega_I}{\partial \rho}+\mu=0,
\end{equation}
and 
\begin{equation}
\label{eq:EL_psi}
\frac{d}{dz} \frac{\partial \omega_{II}}{\partial \dot{\rho}_s}-\frac{\partial \omega_{II}}{\partial \rho_s}=0.
\end{equation}
Note that the condition of $\rho$ being $z$-independent requires that the profiles $\rho_n(z)$ and $\rho_s(z)$ are connected; a change in one of them leads to a change in the other. We stress that the above arguments are \textit{not} dependent on the actual functional form of $\omega$; they rely simply on the assumption that the overall density of the fluid inside the film does not change. 

Within the $\Psi$ theory the type of phase transition in helium films from
helium I to helium II depends crucially on the value of the parameter $M$ \cite{GS76,GS82}. For $M < 1$ this transition, as in bulk helium, is continuous, while
for $M > 1$ the transition in a film is first order. The value
$M = 1$ corresponds to a tricritical point. Thus, we use $M<1$ in our calculations. Obviously, the simplest case has $M=0$. 

Keeping in mind \eq{eq:omegaII}  and \eq{eq:EL_psi} for the function $\Psi$, one obtains the equation
\begin{equation}
\label{eq:psi_eq}
\frac{\hbar^2}{2m} \ddot{\Psi}=\Psi \frac{\partial \omega_{II}}{\partial |\Psi|^2},
\end{equation}
or, in terms of the reduced variable $\psi$
\begin{equation}
\label{eq:psi_red_eq}
\Psi_{s,e0}^2 \, \frac{\hbar^2}{2m} \ddot{\psi}=\psi \frac{\partial \omega_{II}}{\partial |\psi|^2},
\end{equation}
Introducing, as in \cite{GS76,GS82}, the scaled spatial variable
\begin{equation}
\label{eq:zeta_def}
\zeta_0= z/\xi_0,
\end{equation}
where, see Eq. (23) in \cite{GS82}, for $\xi_0$ one has
\begin{equation}
\label{eq:xi0}
\xi_0=\frac{\hbar \Psi_{s,e0}}{\sqrt{2m T_\lambda \Delta C_\mu}}=\frac{\hbar}{m}\sqrt{\frac{\rho_{s,0}}{2 T_\lambda \Delta C_\mu}}\simeq 1.63 \times 10^{-8} {\rm cm}
\end{equation}
 with $\xi_0$ being the amplitude of the correlation function above the $\lambda$ point for the version of the theory
with $M = 0$, one can write the equation for the dimensionless function $\psi$ in the form 
\begin{equation}
\label{eq:psi_red_eq_expl}
\ddot{\psi}=\dfrac{3}{3+M} \psi \left[-\tau |\tau|^{1/3} + (1-M) |\tau|^{2/3} \left|\psi\right|^2 + M\left|\psi\right|^4 \right].
\end{equation}
Here the differentiation is to be understood with respect to the scaled variable $\zeta_0$. \eq{eq:psi_red_eq_expl} is the main equation within the $\Psi$ theory one deals with.

The proper boundary conditions are
\begin{equation}
\label{eq:bc_spec}
\psi(0)=0, \psi(L)=0,
\end{equation}
but so that $\lim_{L\to \infty} \psi(L/2)=\psi(\infty)=\psi_e=\tau^{1/3}$.

\section{The behavior of the Casimir force}

It is convenient to introduce the variables
\begin{equation}
\label{eq:xi_t_psi_theory}
\xi_\tau=\sqrt{\dfrac{3+M}{3}}\,\xi_0\, |\tau|^{-2/3}, \quad \mbox{and} \quad \phi=\psi |\tau|^{-1/3},
\end{equation}
where $\xi_0$ is given by \eq{eq:xi0}. Then \eq{eq:psi_red_eq_expl} becomes 
\begin{equation}
\label{eq:phi}
\ddot{\phi}=\phi \left[-\sign(\tau) + (1-M) \left|\phi\right|^2 + M\left|\phi\right|^4 \right],
\end{equation}
where the derivative is taken with respect to $\zeta_\tau$,
\begin{equation}
\label{eq:zeta_tau}
\zeta_\tau \equiv \frac{z}{\xi_\tau}=\frac{z}{L}\frac{L}{\xi_\tau}=\zeta_L x_\tau, 
\end{equation}
with 
\begin{equation}
\label{eq:zeta_L_x_tau}
\zeta_L=\frac{z}{L} \quad \mbox{and} \quad  x_\tau=\frac{L}{\xi_\tau}.
\end{equation}
Note that since $\xi_\tau$ depends on $M$, see \eq{eq:xi_t_psi_theory}, the scaling variable $x_\tau$, see \eq{eq:zeta_L_x_tau}, is also $M$-dependent. Note also that, in contrast to commonly utilized notations, $\tau>0$ corresponds to $T<T_\lambda$. Obviously, in equilibrium bulk helium, when $\ddot \phi=0$, one has  $\phi \equiv \phi_b$ with $\phi_b=1$ for $T\le T_\lambda$, and $\phi_b=0$ for $T>T_\lambda$. 

Now we turn to the solution of this equation in a system with a film geometry.

Multiplying  (\ref{eq:phi}) by $\dot{\phi}$ and integrating, we obtain
\begin{equation}
\label{eq:invariant}
\dot{\phi}^2+\sign(\tau) \phi^2 - \dfrac{1}{2}(1-M) \phi^4 - \dfrac{1}{3}M\phi^6=p,
\end{equation}
where $p$ is a quantity that is $z$-independent. One should also note that $\dot{\phi}=0$ at $z=L/2$. Let us denote $\phi(z=L/2)=\phi_0$. Then one has 
\begin{equation}
\label{eq:p}
p=\sign(\tau) \phi_0^2 - \dfrac{1}{2}(1-M) \phi_0^4 - \dfrac{1}{3}M\phi_0^6.
\end{equation}
Thus, for $\dot{\phi}^2$ one has 
\begin{equation}
\label{eq:phi_dot}
\dot{\phi}^2=\sign(\tau)(\phi_0^2-\phi^2)-\dfrac{1}{2}(1-M) (\phi_0^4-\phi^4)- \dfrac{1}{3}M (\phi_0^6-\phi^6).
\end{equation}
At the boundary we have $\phi(0)=0$. This is the minimum value of $\phi$. 
For $T<T_\lambda$, i.e. $\tau>0$ the derivative is greater than zero for $\zeta_\tau$ in the interval  from $\zeta_\tau=0$ to the middle of the system, where it vanishes when the profile levels off close to its bulk value of $\phi_b=1$.  For $T>T_\lambda$, i.e. $\tau<0$, one finds that $\dot{\phi}^2\le 0$ if $\phi(\zeta_\tau)<\phi_0$ for any value of $\zeta_\tau$.  Keeping in mind the fact that $\phi(0)=0$, we conclude that $\phi(\zeta_\tau)=0$ is the only possible solution in this case.   

Before proceeding to the technical details of the calculations let us note that, according to \eq{eq:first_int}, one has
\begin{equation}\label{p_versus_PL}
p=\dfrac{1}{2} P_L,
\end{equation}
where $P_L$ is the pressure on the boundaries of a system with size $L$, the behavior of which is mathematically described  by the corresponding functional written in terms of the variable $\phi$. In terms of $\phi$ and $x_\tau$,  \eq{eq:II_pot} for $\omega _{\text{II},0}$ becomes 
\begin{eqnarray}\label{eq:omega2viaphi}
\beta \omega _{\text{II},0} &=& L^{-3} \sqrt{\frac{3+M}{3}} \beta T_{\lambda } \text{$\Delta $C}_{\mu }\xi _0^3\\ && \times \; x_{\tau }^3
\left(-\sign(x_{\tau }) \phi ^2+\frac{1}{2} (1-M) \phi ^4+\frac{1}{3}M \phi
^6 \right) .\nonumber
\end{eqnarray}

The equation for the profile $\phi(\zeta_\tau)$, $\tau>0$, reads
\begin{widetext}
\begin{equation}
\label{eq:profile}
\zeta_\tau=\int_{0}^{\phi (\zeta_\tau)} \frac{d\phi}{\sqrt{(\phi_0^2-\phi^2)-\dfrac{1}{2}(1-M) (\phi_0^4-\phi^4)- \dfrac{1}{3}M (\phi_0^6-\phi^6)}},
\end{equation}
complemented by the equation that determines $\phi_0$
\begin{equation}
\label{eq:phi_not}
\frac{1}{2}\dfrac{L}{\xi_\tau}=\int_{0}^{\phi_0} \frac{d\phi}{\sqrt{(\phi_0^2-\phi^2)-\dfrac{1}{2}(1-M) (\phi_0^4-\phi^4)- \dfrac{1}{3}M (\phi_0^6-\phi^6)}}.
\end{equation}
Introducing the variable $\phi=y \phi_0$, and after that performing the change of variables from  $y^2\to y$ the above equation becomes 
\begin{equation}
\label{eq:phi_not_y}
x_\tau \equiv \dfrac{L}{\xi_\tau}= \int_{0}^{1} \frac{dy}{\sqrt{y\left(1-y\right)\left[1-\dfrac{1}{2}(1-M)\phi_0^2 \left(1+y\right)- \dfrac{1}{3}M \phi_0^4\left(1+y^2+y \right)\right]}}.
\end{equation}
\end{widetext}
The integral on the right-hand side of \eq{eq:phi_not_y}  leads naturally to expressions in involving elliptic functions. 

\subsection{The case $M=0$}
In this case the expression for $p$ becomes 
\begin{equation}
\label{eq:p}
p=\sign(\tau) \phi_0^2 - \dfrac{1}{2} \phi_0^4,
\end{equation}
while for $x_\tau$, from \eq{eq:phi_not_y}, one has
\begin{equation}
\label{eq:xt}
x_{\tau } = \int_0^1 \frac{1}{\sqrt{y (1-y) \left(\text{sgn}(\tau )-\frac{1}{2} \phi _0^2
		(1+y)\right)}} \, dy.
\end{equation}

From \eq{eq:xt} it is clear that $x_\tau$ is a well defined quantity only for $0\le\phi_0<1$. It is also easy to check that $x_\tau$ is a monotonically increasing function of $\phi_0$, with $x_\tau(\phi_0=0)=\pi$. The last implies that one will have a non-zero solution for $\phi_0$ and, therefore, for $\phi(\zeta_\tau)$ for $x_\tau>\pi$. Let us also note that from \eq{eq:p} one concludes $0 \le p<1/2$.  

Taking the integral in \eq{eq:xt}, when $0\le\phi_0<1$ one derives
\begin{eqnarray}
\label{eq:xt_tau_pos}
x_{\tau }= \frac{2 \sqrt{2} }{\sqrt{2-\phi_0^2}}K\left(\sqrt{\frac{\phi _0^2}{2-\phi_0^2}}\right).
\end{eqnarray}
It is easy to check that the right-hand side of \eq{eq:xt_tau_pos} is a monotonically increasing function of $\phi_0$. Thus, one can uniquely invert this equation, thereby determining  $\phi_0(x_\tau)$. 

As noted previously, when  $\phi_0\ge\phi(\zeta_\tau)$,  one is directly led to the conclusion that $\phi(\zeta_\tau)=0$ is the only allowed solution of \eq{eq:phi_dot}.

\begin{figure}[h!]
	\includegraphics[width=\columnwidth]{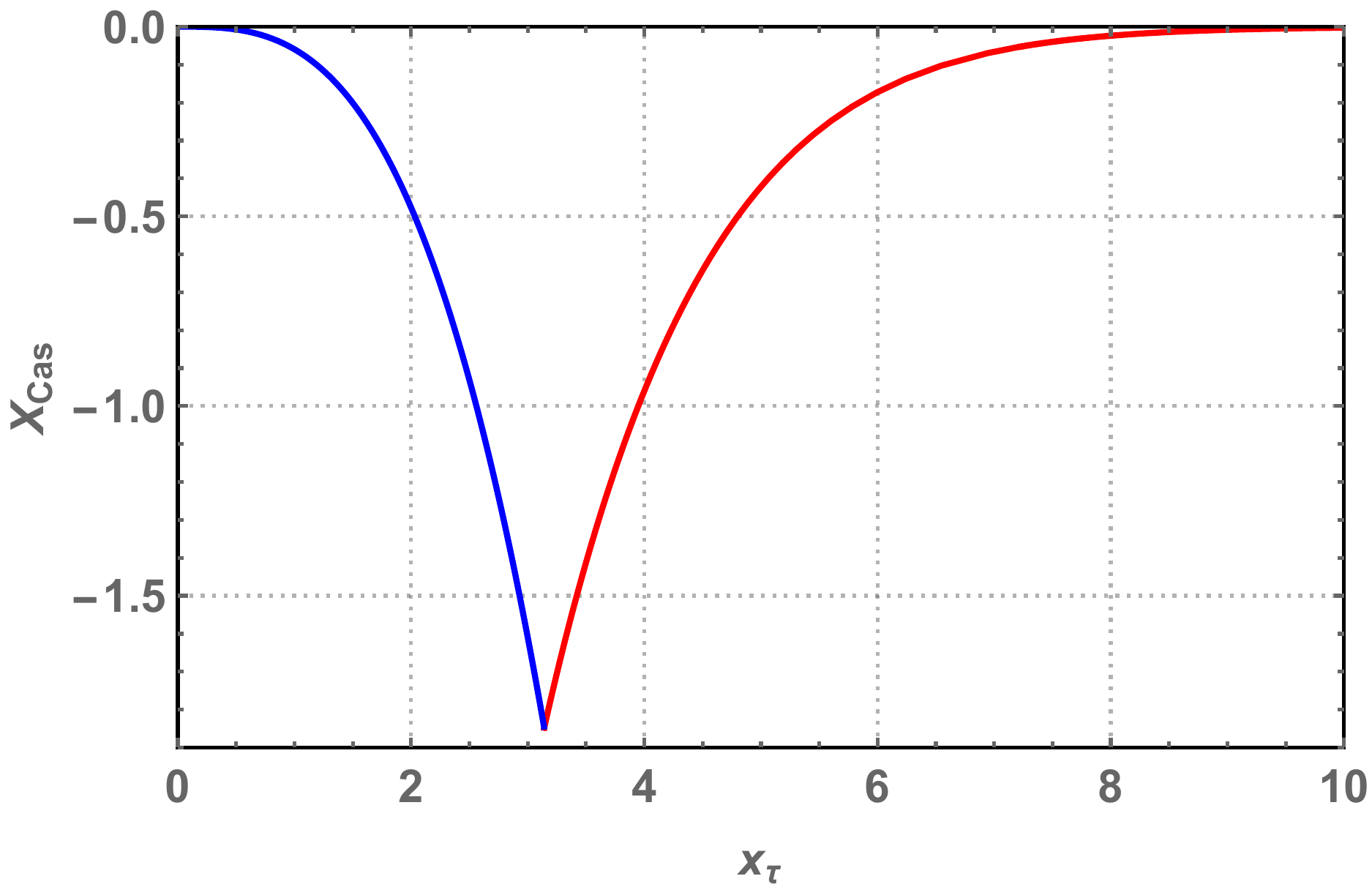}
	\caption{
		The behavior of the Casimir force within the $\Psi$ theory when $M=0$.  } 
	\label{fig:PsiSRCF}
\end{figure}

Summarizing the information from the two above sub-cases $\tau>0$ and $\tau<0$, and performing the corresponding numerical evaluations for the behavior of the Casimir force in the  case $M=0$ one obtains the result shown in Fig. \ref{fig:PsiSRCF}.
In order to obtain this curve we make use of the first integral \eq{eq:p}, its relation to the pressure in the finite system \eq{p_versus_PL}, the corresponding easily obtainable expression for the bulk pressure, as well as the relation \eq{eq:xt} between $\phi_0$ and $x_t$, and, finally \eq{eq:omega2viaphi}, which becomes
\begin{eqnarray}\label{eq:Xcas_how}
\beta F_{\rm Cas}(T,L)&=&\dfrac{1}{2} \beta T_{\lambda } \Delta C_{\mu }\xi _0^3\; x_\tau^3 \left[p(\phi_0(x_t))-1/2\right] L^{-3} \nonumber \\
&\simeq& 0.119 \; x_\tau^3 \left[p(\phi_0(x_t))-1/2\right] L^{-3}. 
\end{eqnarray}
The evaluation of the above expression leads us to the curve displayed in Fig. \ref{fig:PsiSRCF} with
\begin{equation}\label{eq:XCas}
X_{\rm Cas}(x_\tau)=0.119 \; x_\tau^3 \left[p(\phi_0(x_t))-1/2\right].
\end{equation}
\begin{figure}[h!]
	\includegraphics[width=\columnwidth]{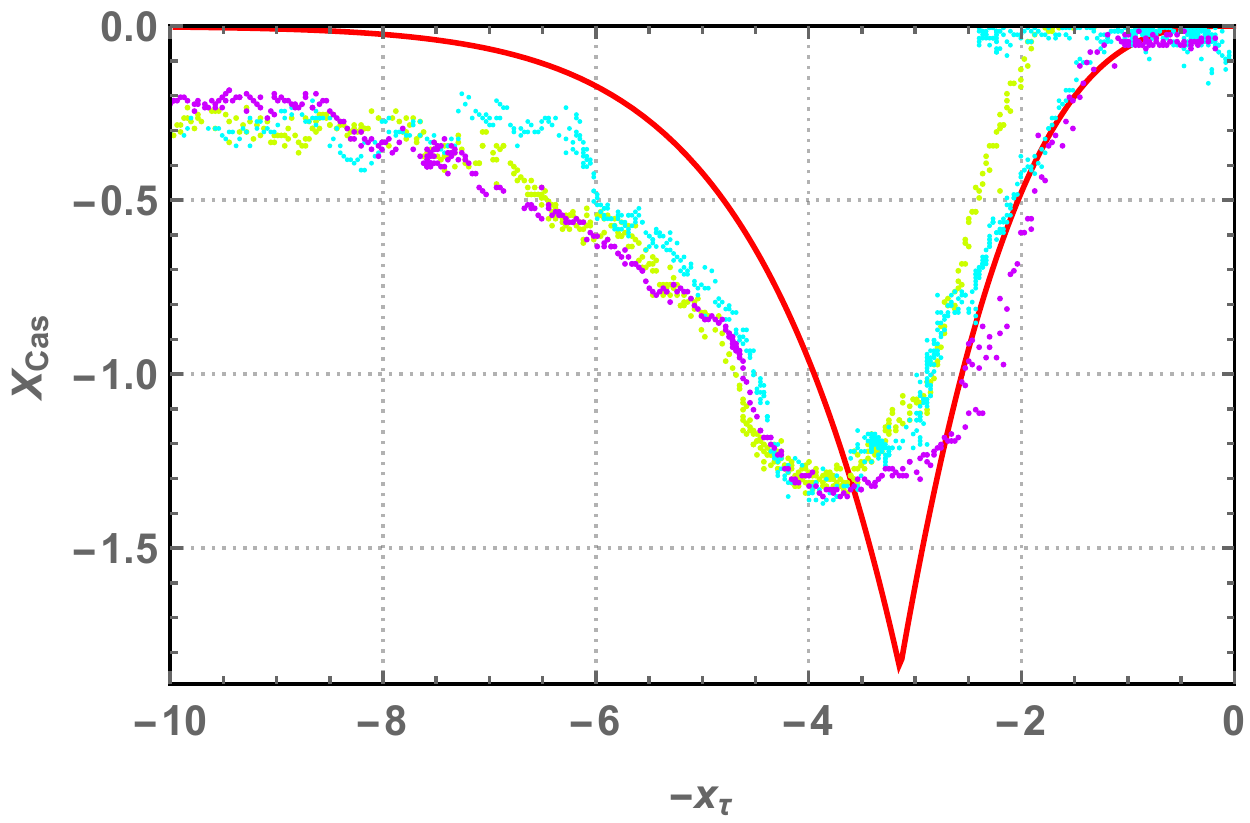}
	\caption{
		A comparison of the experimental data of the Casimir force, the scattered curves,  reported in \cite{GSGC2006}  with the prediction of the $\psi$ theory when $M=0$, the solid curve.
		The position of the minimum within the $\Psi$-theory is at $x_\tau=\pi$, while the experiment delivers $x_\tau=3.8$. The minimal value of the force is $-1.848$ within the theory, while within the experiment it is $-1.30$.} 
	\label{fig:PsiSRCF_comp}
\end{figure}

Solving \eq{eq:profile}, for the order parameter profile $\phi(\zeta_\tau)$ in the case $M=0$ one has
\begin{equation}\label{eq:profileM0}
\phi(\zeta_\tau)=\phi _0 \; \text{sn}\left(\zeta_\tau  \sqrt{1-\frac{\phi _0^2}{2}}\Bigg|\sqrt{\frac{\phi
		_0^2}{2-\phi _0^2}}\right),
\end{equation}
where $\phi_0$, as a function of $x_\tau$, is to be determined from \eq{eq:xt_tau_pos}. Here ${\rm sn}$ is the Jacobi elliptic function ${\rm sn}(u|m)$ \cite{AS}. 

Fig. \ref{fig:PsiSRCF_comp} presents a comparison of the experimental determination \cite{GSGC2006} of the Casimir force  with the prediction of the $\psi$ theory with $M=0$. When transferring the experimental data of \cite{GSGC2006}, given in terms of $(T/T_\lambda-1)L^{1/\nu}$ to the variable $L/(\xi_0 |t|^{1/\nu})$ we have used the value of $\nu$ given in \eq{eq:crit_exp}, and the data of $\xi_0=1.2$ $\AA$ reported in \cite{IP74}. We observe that while the position of the minimum within the $\Psi$-theory is at $x_\tau=\pi$, the experiment yields $x_\tau=3.8$. The minimal value of the scaling function of the force is $-1.848$ within the theory, and $-1.30$ in the experiment.

\subsection{The case $0<M<1$}

From \eq{eq:phi_not_y} it is easy to check that $x_\tau$ is a monotonically increasing function of $\phi_0$, with $x_\tau(\phi_0=0, M)=\pi$. The last implies that one will have a non-zero unique solution for $\phi_0$ and, therefore, for $\phi(\zeta_\tau)$ only when $x_\tau>\pi$. The above statements are valid for {\it any} value of $0\le M<1$. 

The Casimir force is now given by the expression
\begin{eqnarray}\label{eq:Xcas_how_M}
\beta F_{\rm Cas}(T,L;M)&=&\dfrac{1}{2} \beta T_{\lambda } \Delta C_{\mu } \xi _0^3\;\sqrt{\dfrac{3+M}{3}}\; x_\tau^3(M) \\
&& \times  \left[p(\phi_0(x_t,M), M)-\dfrac{3+M}{6}\right] L^{-3} \nonumber \\
&\simeq& 0.119\sqrt{\dfrac{3+M}{3}} \; x_\tau^3 \left[p-\dfrac{3+M}{6}\right] L^{-3} \nonumber. 
\end{eqnarray}
Evaluating the integral in \eq{eq:phi_not_y}, for $\phi_0$ one obtains the equation
\begin{equation}
\label{eq:phi_not_y1}
x_\tau \equiv \dfrac{L}{\xi_\tau}= \frac{4 \sqrt{3}}{b} K(k)
\end{equation}
where $K(k)$ is the complete elliptic integral of elliptic modulus 
\begin{equation}
k=\sqrt{ 2 \sqrt{3} a}\; \frac{\phi_0 }{b}
\end{equation} 
with
\begin{eqnarray}
a &=&\sqrt{(3+M+2M\phi_0^2) (1+M(3-2\phi_0^2)) }, \nonumber \\
b &=&\sqrt{12-6 M\phi_0^4+\phi_0^2 ( \sqrt{3} a-9(1-M))}. \nonumber
\end{eqnarray}

Solving this equation for $\phi_0$, one finds $\phi_0=\phi_0(x_\tau, M)$, where $x_\tau$ also depends on $M$ via \eq{eq:xi_t_psi_theory} and \eq{eq:zeta_L_x_tau}. 

\begin{figure}[h!]
	\includegraphics[width=\columnwidth]{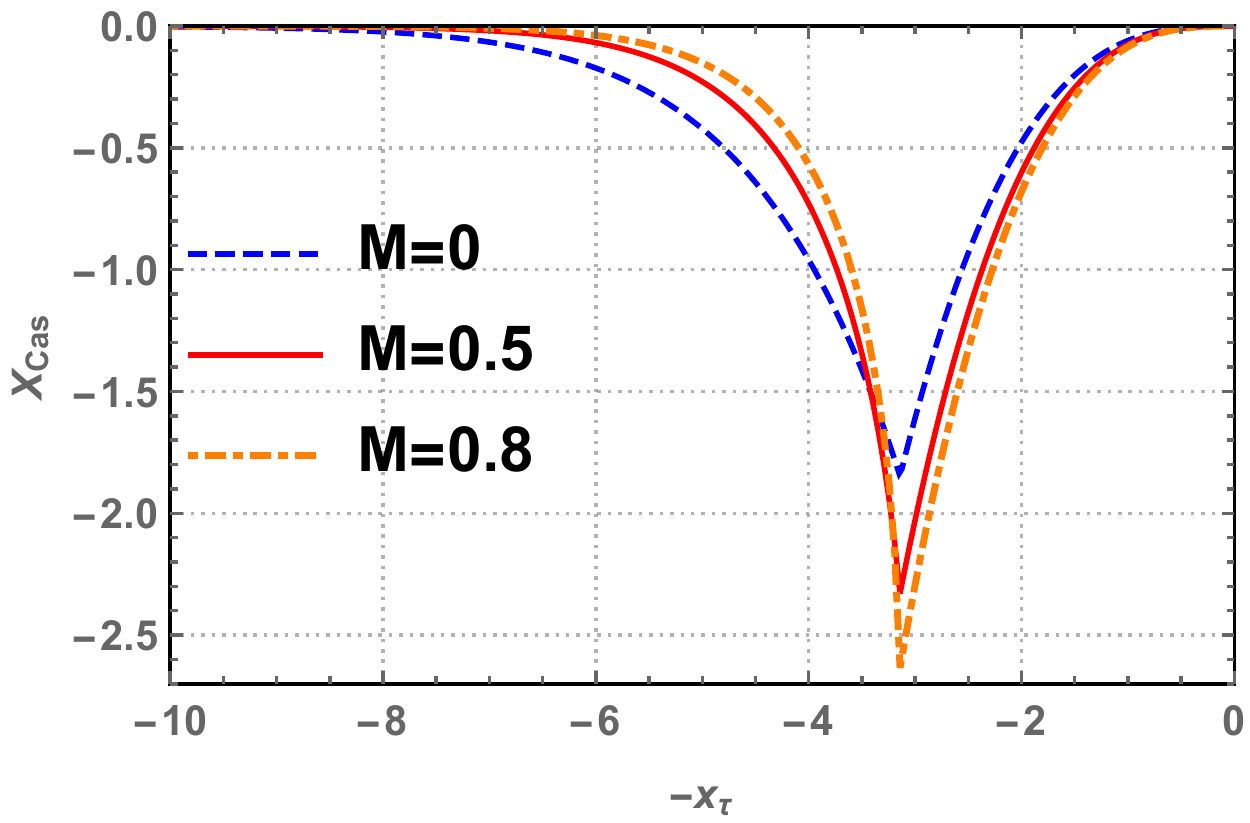}
	\caption{
		The behavior of the Casimir force within the $\psi$ theory when $M=0$, $M=0.5$ and $M=0.8$. The minimal value of the force gets deeper with increase of M.} 
	\label{fig:PsiSRCFdiffM}
\end{figure}

A comparison of the Casimir force calculated in the  case $M=0$,  $M=0.5$ and $M=0.8$ is shown in Fig. \ref{fig:PsiSRCFdiffM}. The position of the minimum stays unchanged but its absolute value increases with increasing $M$. Thus, the $M=0$ theory coincides most closely with the experimentally reported data. 

The order parameter profile can be also obtained in an explicit form for the case $0\le M<1$. Solving \eq{eq:profile}, for the order parameter profile $\phi(\zeta_\tau|M)$ one has
\begin{equation}\label{eq:profileM_not_0}
\phi(\zeta_\tau|M)=\frac{\sqrt{c} \, \phi _0  \, \mathrm{sn}\left(\frac{b}{2 \sqrt{3}} \zeta_{\tau} | k \right)}{\sqrt{1+c-\mathrm{sn} \left(\frac{b}{2 \sqrt{3}}\, \zeta_{\tau} |k \right)^2}},
\end{equation}
where
\begin{equation}
c=\frac{\sqrt{3} a+M \left(2 \phi_0 ^2 -3\right)+3}{4 M \phi_0 ^2}.
\end{equation}
Here $\phi_0$, as a function of $x_\tau$, is to be determined from \eq{eq:phi_not_y1}. Here  ${\rm sn}$ again symbolizes the Jacobi elliptic function ${\rm sn}(u|m)$.

\section{Discussion and concluding remarks}

In the current study we have applied the $\Psi$-theory of Ginzburg and co-authors \cite{GS76,GS82} to evaluate the Casimir force in $^4$He film in equilibrium with its vapor. We have obtained an exact closed form expression for the force within this theory---see \eq{eq:Xcas_how}---when the parameter of the theory $M=0$,  and \eq{eq:Xcas_how_M} for $0\le M<1$. We have found the best agreement between the theory and experiment for $M=0$. The corresponding result for the scaling function of the Casimir force is shown in Fig. \ref{fig:PsiSRCF} and the comparison with the experiment is shown in Fig.  \ref{fig:PsiSRCF_comp}. We conclude that there is reasonably good agreement between this model theory and experiment. One should note, however, some important differences. In the $\Psi$ theory there is a sharp two-dimensional phase transition with long-ranged order below the critical temperature of the finite system, while in the helium system one expects a Kosterlitz-Thouless type transition. This feature  is not captured by the $\Psi$-theory. Also missing are the Goldstone modes and surface wave contributions that appear at low temperatures \cite{ZRK2004}. The overall agreement between the result of the $\Psi$-theory and the experiment, shown in Fig. 	\ref{fig:PsiSRCF_comp}  is, however, much better than is provided by mean-field theory \cite{ZSRKC2007}; there is no need to fit any parameter in $\Psi$ theory in order to achieve this agreement.  
\acknowledgements{We are indebted to the authors of \cite{GC99} and \cite{GSGC2006} for providing their experimental data in electronic form. 
	
D. D., V. V. and P. D. gratefully acknowledge the  financial support via contract DN02/8 of Bulgarian NSF. J. R. is pleased to acknowledge support from the NSF through DMR Grant No. 1006128}.

\end{document}